\documentclass[trackchanges,twocolumn]{aastex7}

\begin{document}

\title{Exploring the Frontiers of Cosmic Ray Physics: Perspectives on GRANDProto300 and the GRAND Project}

\author{Kewen Zhang}
\affiliation{
    {Key Laboratory of Dark Matter and Space Astronomy, Purple Mountain Observatory, Chinese Academy of Sciences}, No. 10 Yuanhua Road, Nanjing, 
    China}
\affiliation{School of Astronomy and Space Science, University of Science and Technology of China, Hefei 230026, China}
\email{kwzhang@pmo.ac.cn}

\author{Yi Zhang}
\affiliation{
    {Key Laboratory of Dark Matter and Space Astronomy, Purple Mountain Observatory, Chinese Academy of Sciences}, No. 10 Yuanhua Road, Nanjing, 
    China}
\affiliation{School of Astronomy and Space Science, University of Science and Technology of China, Hefei 230026, China}
\email{zhangyi@pmo.ac.cn}
 
\author{Yi-Qing Guo}
\affiliation{Key Laboratory of Particle Astrophysics, Institute of High Energy Physics, Chinese Academy of Sciences, Beijing 100049, China}
\affiliation{University of Chinese Academy of Sciences, Beijing 100049, China}
\affiliation{Tianfu Cosmic Ray Research Center, Chengdu, Sichuan 610213, China}
\email{guoyq@ihep.ac.cn}

\correspondingauthor{Yi Zhang}
\email{zhangyi@pmo.ac.cn}  

\begin{abstract}
The Giant Radio Array for Neutrino Detection (GRAND) is an envisioned large-scale radio array designed to detect ultra-high-energy cosmic rays (UHECRs, $E > 100$ PeV) and neutrinos. Employing cost-effective antennas distributed across vast areas, GRAND is optimized to observe the rare flux of ultra-high-energy particles with high precision. The GRANDProto300 (GP300) pathfinder array, currently under deployment, targets the $10^{16.5} - 10^{18}$ eV range and is anticipated to achieve approximately 15\% energy resolution and 20g/cm$^2$ $X_{\mathrm{max}}$ precision. This level of precision enables accurate measurements of the fine structure of the energy spectrum, mean logarithmic mass ($\langle \ln A \rangle$), and proton flux within this range. After five years of data collection, the sensitivity for detecting anisotropy could reach $5 \times 10^{-3}$ for energies below $10^{17.1}$ eV. With its substantially larger effective area, GRAND extends these capabilities to the highest energies ($\sim 10^{20}$ eV), offering enhanced statistics and sensitivity for spectral, composition, and anisotropy measurements within one year for UHECRs.
\end{abstract}



\section{Introduction}
Ultra-high-energy cosmic rays (UHECRs), with energies above 100 PeV \citep{coleman2023ultra}, probe the universe’s most extreme environments, offering insights beyond the capabilities of terrestrial accelerators. Studying UHECRs is essential for understanding the mechanisms of cosmic particle acceleration and their propagation across cosmological distances. Despite significant progress, several key questions remain unanswered, particularly regarding their origin, composition, and intergalactic propagation. Addressing these questions requires
precise measurements of observables such as the energy spectrum, mass composition, and large-scale anisotropy.

The UHECRs energy spectrum provides a coherent view of the transition from Galactic to extragalactic cosmic rays.
This spectrum has revealed several distinct features: the knee at approximately 5 $\times 10^{15}$ eV \citep{fowler2001measurement}, the low-energy ankle just above 10$^{16}$ eV \citep{novotny2022energy}, the second knee near 10$^{17}$ eV \citep{candia2003cosmic}, the ankle around 5 $\times 10^{18}$ eV \citep{guido2022combined}, the recently identified instep at about 10$^{19}$ eV \citep{aab2020measurement}, and the suppression beginning near 5 $\times 10^{19}$ eV, often associated with the Greisen-Zatsepin-Kuzmin (GZK) cutoff \citep{greisen1966end, zatsepin1966upper}. These features encode vital information about the sources, their cosmological
distribution, and energy losses due to interactions with the cosmic microwave
background.

While anisotropy measurements support an extragalactic origin for UHECRs above $\sim$ 8 EeV, the energy 
 \citep{pierre2017observation} the energy range between 100 PeV and 10 EeV is poorly constrained  \citep{coleman2023ultra}. This range is critical, as it marks the transition from Galactic to extragalactic sources. High-statistics measurements in this region are essential to distinguish between competing models and reduce cross-experiment systematic errors. Notably, mass composition measurements from 100 PeV to 1 EeV provide critical information complementary to spectral features, probing the transition from Galactic to extragalactic origins.
The primary observable, X$_{\mathrm{max}}$—the atmospheric depth at which an air shower reaches its maximum—is sensitive to the initiating particle’s mass. However, inferring composition, typically expressed as the mean logarithmic mass ($\langle \ln A \rangle$), requires comparison to air shower simulations, rendering results highly dependent on hadronic interaction models. Measurements of $\langle \ln A \rangle$ reveal energy-dependent trends, but inconsistencies in normalization and slope among fluorescence-based experiments, such as TALE, Tunka-133, and Yakutsk, highlight persistent systematic uncertainties\citep{coleman2023ultra}.

Above 1 EeV, results from Pierre Auger Observatory (PAO) and Yakutsk suggest a gradual shift from light to heavier primaries with increasing energy, as protons and helium become depleted. While TA results suggest a lighter composition, they remain statistically compatible with PAO within current uncertainties. Elongation rate studies from multiple observatories across both hemispheres support a transition around 3 EeV \citep{coleman2023ultra}, consistent with a shift toward heavier nuclei. However, beyond the suppression (E $> 10^{19.6}$ eV), low event statistics limit robust composition determination.

To rigorously address these open questions, particularly in the 100 PeV to multi-EeV range, further measurements with enhanced precision, wide-field- of-view sky coverage, and complementary detection techniques are necessary for cross-validation and to mitigate systematic biases.

In this context, radio detection of extensive air showers has emerged as a powerful and increasingly mature technique. Compared to traditional fluorescence and surface detectors, radio arrays offer several advantages: they operate with a duty cycle close to 100\%, measures sole electromagnetic component of air showers, and can achieve high precision in energy \citep{aab2016energy, Schlüter_2023} and composition reconstruction \citep{buitink2014method,halim2023demonstrating}. Modern digital radio arrays can also cover vast areas at relatively low cost, making them well suited for large-scale UHECRs observatories. Over the past decade, significant progress has been made in modeling radio emission mechanisms, reconstructing shower parameters such as X$_{\mathrm{max}}$, and validating results against established techniques. These advances position radio detection as a complementary—and in some cases competitive—approach for future cosmic-ray experiments.

One of the most ambitious efforts in this direction is the Giant Radio Array for Neutrino Detection (GRAND), which aims to leverage the potential of large-scale radio arrays to study UHECRs, neutrinos, and gamma rays with unprecedented sensitivity.

This paper is organized as follows. Section 2 provides an overview of the GRAND and its pathfinder array, GRANDProto300 (GP300), detailing the detection strategy and expected performance. In Section 3, we present projections for key observables—namely the energy spectrum, proton spectrum, mean logarithmic mass ($\langle \ln A \rangle$), and large-scale anisotropy—based on the estimated sensitivities and reconstruction resolutions of GP300 and GRAND. Finally, Section 4 summarizes the expectations and discusses the prospects of large-scale radio detection for advancing our understanding of UHECRs.

\section{GRAND and GP300}

The Giant Radio Array for Neutrino Detection (GRAND) represents a next-generation experiment designed to address the challenge of detecting UHECRs and neutrinos through a large- scale, radio-based methodology. Operating within the 50–200 MHz frequency band, GRAND will consist of up to 20 independent sub-arrays, each containing 10,000 antennas that together cover an area of 10,000 km$^2$. These autonomous detectors are intended to capture the extensive radio footprints of inclined air showers, which can span tens of kilometers. This modular configuration allows GRAND to efficiently gather high-statistics cosmic-ray data, attain sensitivity to the low fluxes of ultra-high-energy neutrinos and gamma rays, and achieve excellent angular resolution through precise reconstruction of shower geometry.

Given the unprecedented scale of this array, deploying a prototype is crucial for validating the detection concept. To this end, the GP300 pathfinder is being installed, comprising 300 antennas distributed over an area ranging from 100 to 300 km$^2$. GP300 focuses on the detection of near-horizontal air showers in the energy range of 10$^{16.5}$ to 10$^{18}$eV. Its objectives include demonstrating the feasibility of autonomous radio detection of UHECRs and refining reconstruction techniques in preparation for the complete GRAND array.

\begin{figure}
    \centering
    \includegraphics[width=0.9\linewidth]{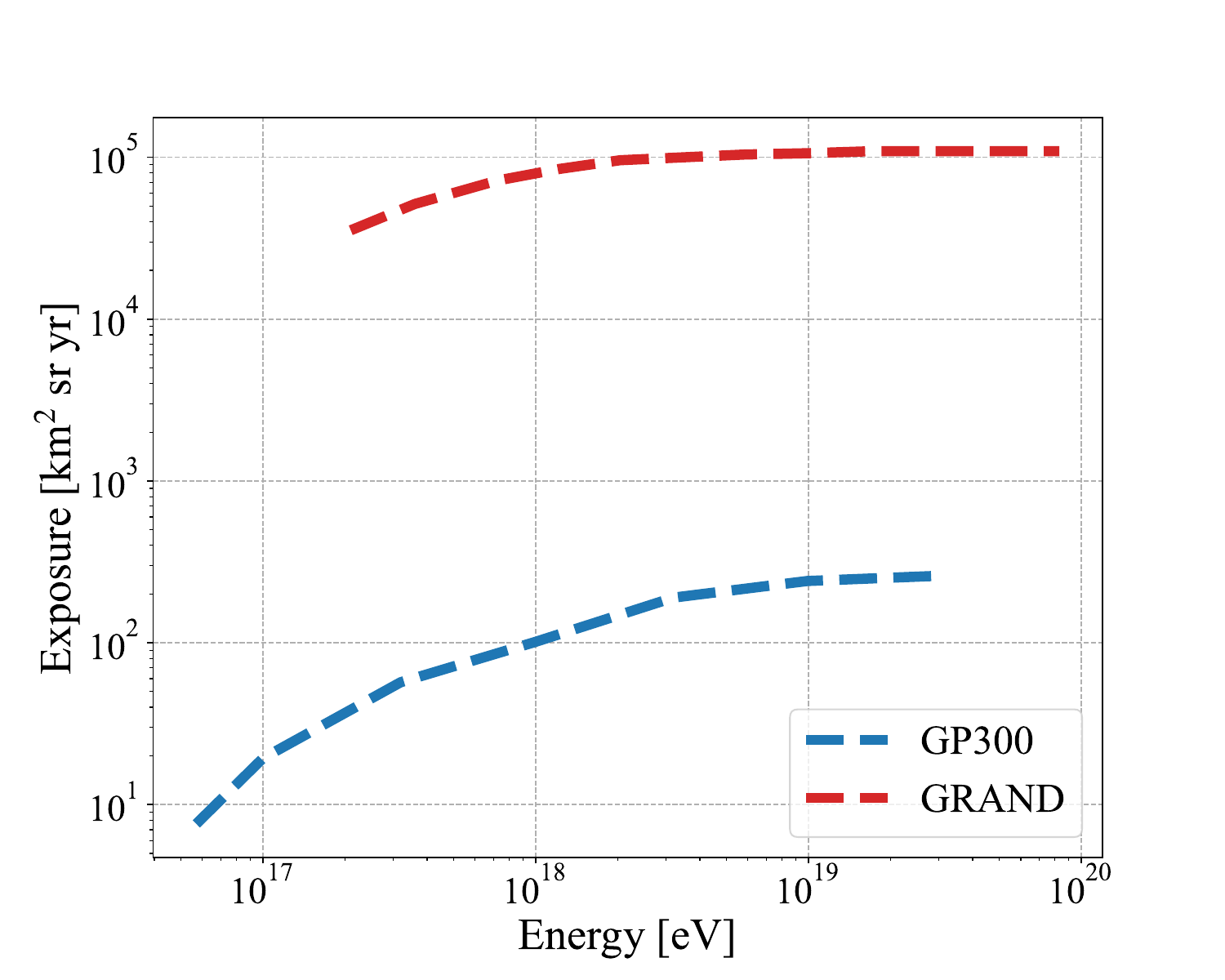}
    \caption{Estimated yearly exposure of GP300 (blue) \citep{alvarez2020giant} and GRAND (red) \citep{coleman2023ultra}, assuming an aggressive detection threshold of 30$\mathrm{\mu}$V, triggered by coincident signals in at least five antennas.}
    \label{fig:exposure}
\end{figure}

Figure \ref{fig:exposure} illustrates the estimated yearly exposure of UHECRs for both GP300 and GRAND. GP300 is anticipated to record approximately 2.5$\times 10^6$ cosmic ray events with energies above 30 PeV per year \citep{alvarez2020giant}In contrast, GRAND aims for an event rate approximately three orders of magnitude higher, extending to the highest energies\citep{coleman2023ultra}.

The GP300 site is located near Dunhuang, China, in a remote desert region with minimal human radio interference. The array employs bow-tie-shaped antennas with three polarization channels, providing broader sky coverage and improved detection capabilities across a wider range of arrival directions.

\section{Expected Performance}

In this section, we present the estimated performance of the GP300 and GRAND experiments in reconstructing key observables of UHECRs, including the energy spectrum, mass composition ($\langle \ln A \rangle$), large-scale anisotropy, and the proton spectrum. These performance estimates are based on detector resolutions and anticipated event statistics derived from simulations of air showers and instrumental responses from \citep{alvarez2020giant}. While GP300 serves as a pathfinder array focusing on horizontal air showers in the $10^{16.5}–10^{18}$ eV range, GRAND, with its much larger aperture, is designed to extend these measurements to higher energies with greater precision. Together, these complementary setups are expected to significantly enhance our understanding of the transition from Galactic to extragalactic cosmic rays and aid in disentangling the underlying sources and acceleration mechanisms.

\subsection{All particle energy spectrum}

The energy resolution of radio detection experiments generally depends on energy, with current radio arrays achieving a resolution of approximately 15\% \citep{alvarez2020giant}. We adopt this value as a conservative and straightforward estimate for our energy spectrum reconstruction. However, it is noteworthy that simulation studies suggest resolutions as high as 5\% may be attainable with optimized reconstruction techniques \citep{Schlüter_2023}.

Various models have been proposed to explain the observed UHECRs spectrum and composition. Karakula and Tkaczyk \citep{karakula1985formation} suggested that interactions between nuclei and background UV photons at the source can shape both spectrum and composition via photodisintegration and photomeson production. More recently, \citep{das2019modeling} introduced a two-population model involving proton-dominated and heavy-nuclei-dominated extragalactic sources, which offers an improved fit to PAO data compared to single-population scenarios. The energy spectrum used in this study is based on measurements above the PeV scale and is modeled through a global fit that incorporates multiple nuclear populations with rigidity-dependent cutoffs. The parameters are derived from the second fit model with four populations, as described by Gaisser \citep{gaisser2013cosmic}, which assumes harder spectral indices and a lower cutoff for the Galactic component ($\mathrm{R_c}$ = 120 TV). This model captures the observed features from below the knee to the ankle and reproduces the general trend of increasing mass with energy. It also facilitates the interpretation of the ankle as a transition between Galactic and extragalactic components.

To estimate the expected energy spectrum observed, we convolve the input flux $\Phi(E)$ with the detector exposure $\mathcal{E}(E)$, across logarithmic energy bins ranging from $10^{16.35}$ to $10^{20.45}$ eV, with a bin width of $\Delta\log_{10}(E/\mathrm{eV}) = 0.3$. The calculation assumes one year of continuous observation at a 100\% duty cycle,  reflecting the operational capabilities expected of radio detection arrays. The expected number of events per bin is given by:

\begin{equation}
    N_{\text{events}}(E) = \Phi(E) \cdot \mathcal{E}(E) \cdot \Delta E \cdot T
    \label{eq:N_evt}
\end{equation}

where $\Delta E$ is the bin width in energy and $T$ is the observation time.

\begin{figure}
    \centering
    \includegraphics[width=0.9\linewidth]{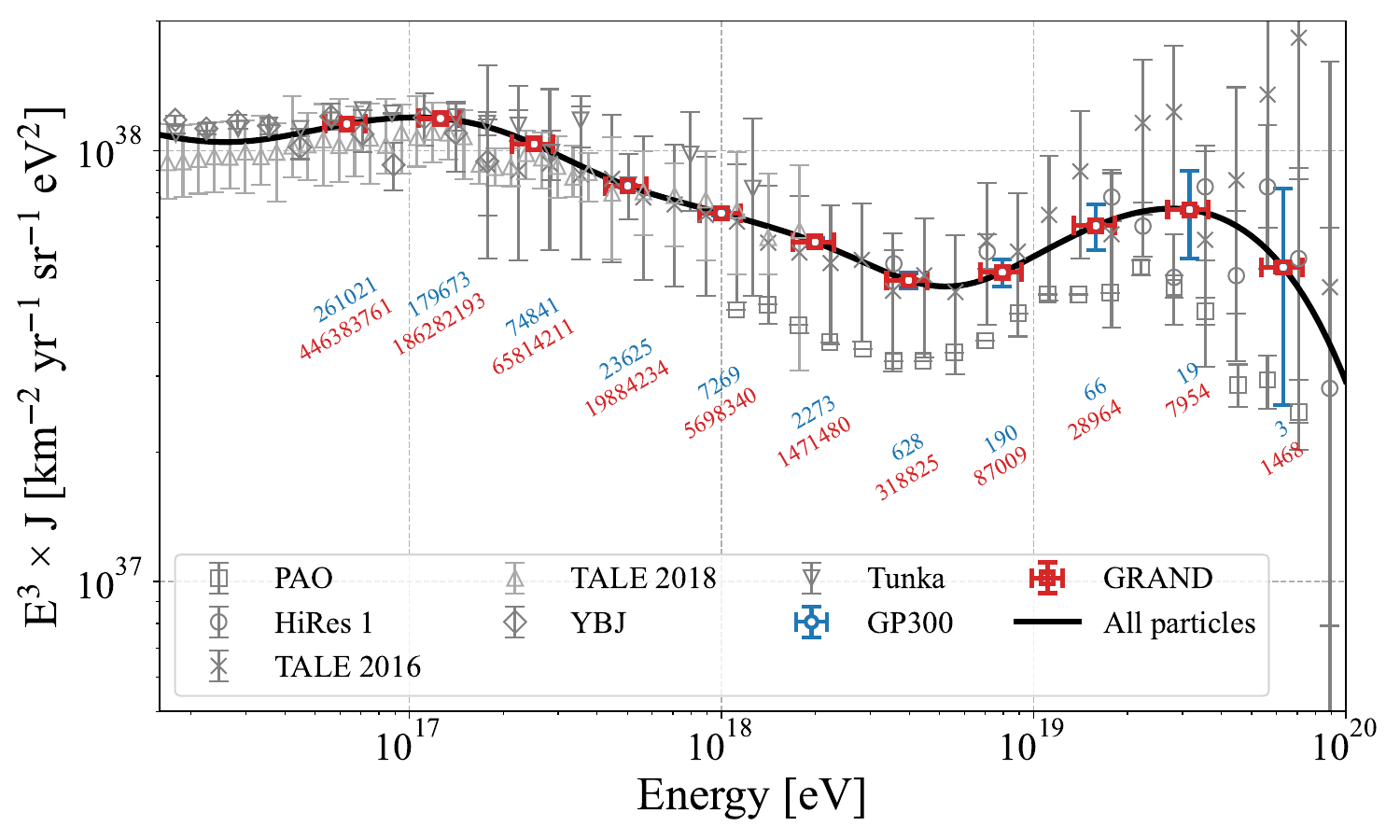}
    \caption{Estimated energy spectrum with statistic uncertainties for GP300 (in blue) and GRAND (in red), along with the corresponding event numbers. Data are compared to current experimental measurements \citep{amenomori2008all, abbasi2008first, budnev2020primary, aab2020features, abbasi2018cosmic, abbasi2016energy}.}
    \label{fig:spectrum}
\end{figure}

Figure \ref{fig:spectrum} illustrates the estimated energy spectra for GP300 and GRAND. Around 100 PeV, GP300 is expected to collect approximately 10$^5$ events, providing robust statistics for spectral measurements in this range. providing robust statistics for spectral measurements in this range. However, the event rate declines markedly above 1 EeV due to the drop in cosmic ray flux and GP300’s limited aperture. In contrast GRAND’s expansive detection area ensures substantial statistics even at the highest energies, enabling precise measurements well into the ultra-high-energy regime.

\subsection{$\langle \ln A \rangle$}

A widely used observable to infer the mass composition of cosmic-ray-induced air showers is the atmospheric depth of the shower maximum, $X_{\mathrm{max}}$, which systematically varies with the mass of the primary particle. This dependence can be described by the generalized Heitler model \citep{matthews2005heitler}, which expresses the mean $X_{\mathrm{max}}$ as:

\begin{equation}
\langle X_{\mathrm{max}} \rangle = X_0 + D \log_{10}\left(\frac{E}{E_0 A}\right),
\label{eq:heitler_mod}
\end{equation}

where $E$ is the primary energy, $A$ the mass number, and $X_0$ is the mean depth of proton showers at energy $E_0$, and $D$ is the elongation rate. 
State-of-the-art reconstruction techniques in radio detection currently achieve $X_{\mathrm{max}}$ resolutions of approximately 17 g/cm$^2$ \citep{buitink2014method}, with recent projections for the Square Kilometre Array (SKA)  suggesting possible resolutions as low as 10 g/cm$^2$  \citep{Buitink:2023rso}.

To estimate the mean logarithmic mass 〈ln A〉 and the average depth of the shower maximum ⟨$X_{\mathrm{max}}$⟩ as a function of energy, we simulate an ensemble of air shower events based on a global fit to the cosmic-ray flux and composition model from Gaisser et al. \citep{gaisser2013cosmic}. For each energy bin defined in the energy spectrum analysis, 5,000 events are generated.

Primary masses are drawn from a representative set of nuclei (p, He, C, O, Fe, 50 $<$ Z $<$ 56 and 78 $<$ Z $<$ 82), with probabilities determined from the energy-dependent composition fractions of the model. For each sampled energy and mass, an $X_{\mathrm{max}}$ value is drawn from a generalized Gumbel distribution \citep{2013interpretation}:

\begin{equation}
\mathcal{G} (z) = \frac{1}{\sigma}\frac{\lambda^\lambda}{\Gamma(\lambda)} e^{-\lambda z - \lambda e^{-z}}, \quad z = \frac{x - \mu}{\sigma}
\end{equation}

where $\mu$, $\sigma$, and $\lambda$ are energy- and mass-dependent parameters representing the location, scale, and shape of the distribution, respectively. This formulation captures the spread and skewness of the shower maximum distributions as a function of primary mass and energy. In this work, we adopt the parameter values derived under the QGSJet II hadronic interaction model \citep{ostapchenko2006nonlinear}.

Each event is weighted according to the expected cosmic-ray flux within its energy bin. Since a fixed number of Monte Carlo events (5,000 per bin) are generated uniformly across each bin, we assign each event a weight proportional to the differential flux at the bin center multiplied by the bin width, i.e., $w_i \propto \Phi(E_i) \cdot \Delta E$. This corrects for the mismatch between the uniform sampling and the steeply falling physical energy distribution. The weighted averages of $\ln A$ and $X_{\mathrm{max}}$ are then computed in each energy bin as:

\begin{equation}
\langle \ln A \rangle = \frac{\sum_i w_i \ln A_i}{\sum_i w_i}, \quad
\langle X_{\mathrm{max}} \rangle = \frac{\sum_i w_i X_{\mathrm{max},i}}{\sum_i w_i},
\end{equation}
where $w_i$ is the event weight. This procedure yields the expected evolution of cosmic-ray mass composition and shower maximum with energy, serving as a reference for comparison with experimental data from radio detection.

\begin{figure}
    \centering
    \includegraphics[width=0.9\linewidth]{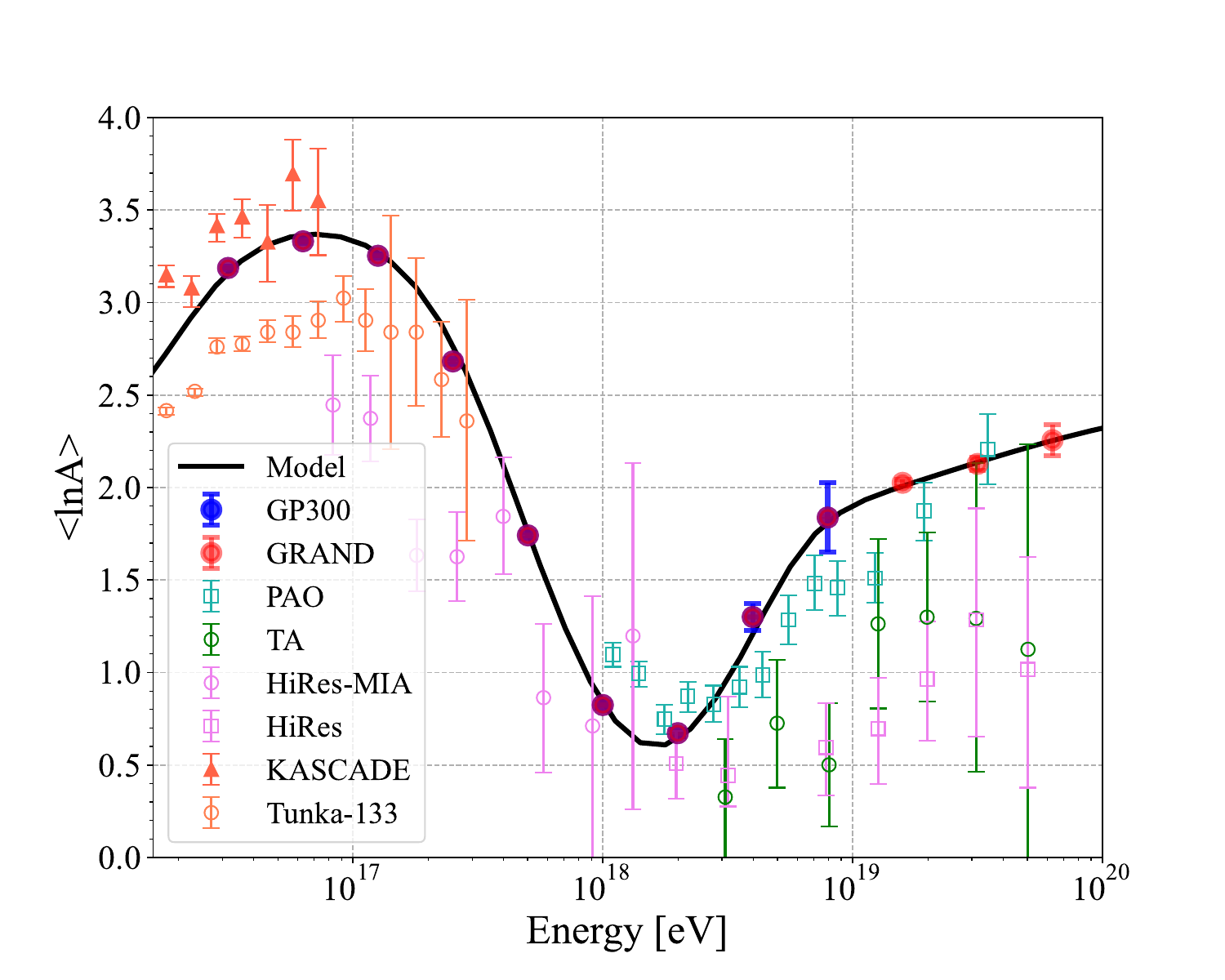}
    \caption{Estimated $\langle \ln A \rangle$ for various interaction models. Dots represent expected values reconstructed under the assumption of QGSJet II, factoring in experimental resolution and statistical uncertainties. The data are compared with current experimental findings from \citep{berezhnev2012tunka, antoni2005kascade, abu2001measurement, abraham2010measurement, abbasi2009measurement, fukushima2015recent}.}
    \label{fig:lnA}
\end{figure}

Figure \ref{fig:lnA} presents the expected evolution of the mean logarithmic mass ⟨ln A⟩ as a function of energy, assuming a radio-based $X_{\mathrm{max}}$ reconstruction resolution of 20 g/cm$^2$ \citep{alvarez2020giant} and a 50\% selection efficiency for high-quality events.

\subsubsection{Proton spectrum}

\begin{figure}
    \centering
    \includegraphics[width=0.9\linewidth]{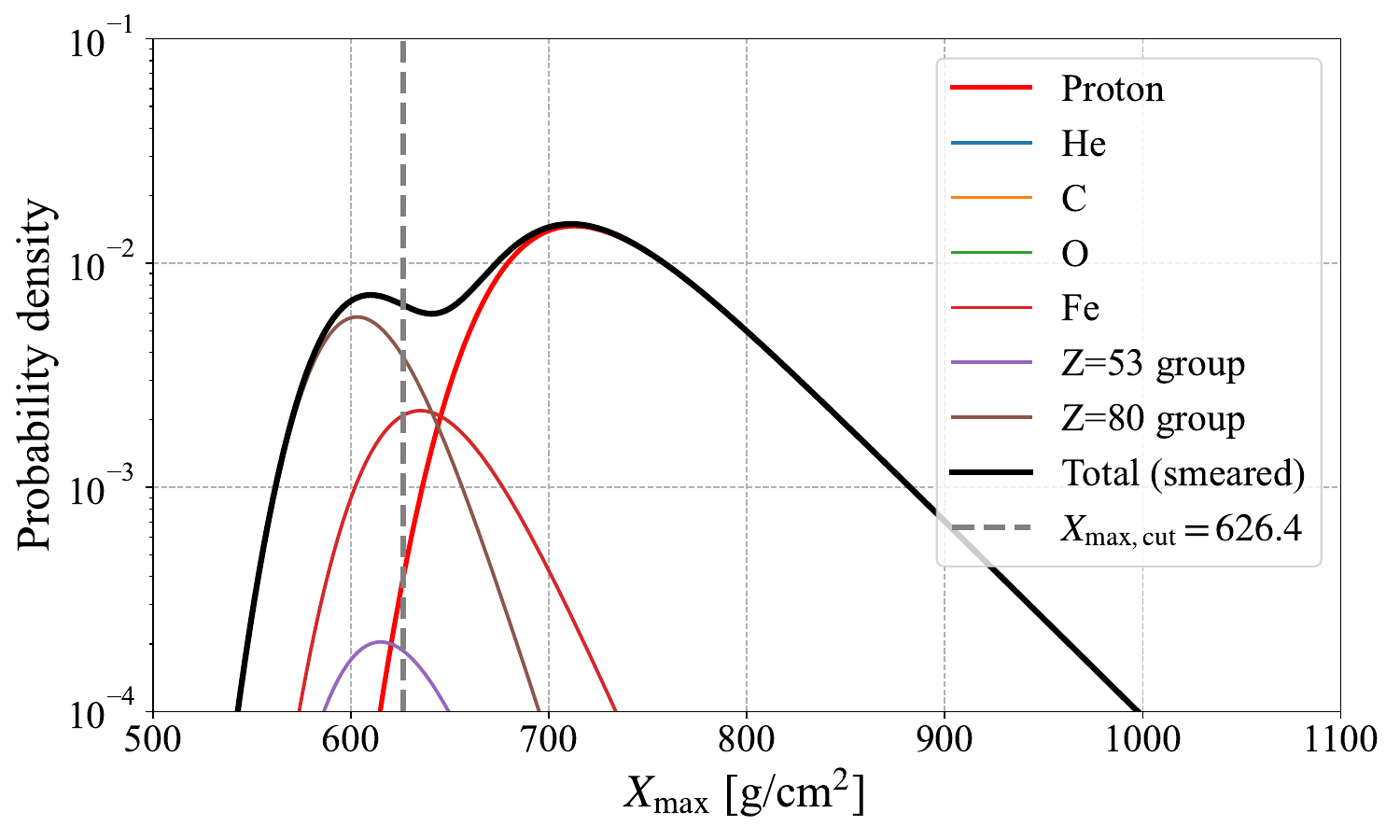}
    \caption{Example distribution of $X_{\mathrm{max}}$ at $E = 10^{18}$ eV with 20 g/cm$^2$ Gaussian smearing. The grey dashed line indicates the $X_{\mathrm{max}}$ cut applied to achieve 90\% proton purity, used to select proton-induced air shower events. According to the model from \citep{gaisser2013cosmic}, the composition at this energy is predominantly proton, with small fraction of iron and heavier nuclei.}
    \label{fig:proton_cut}
\end{figure}

Although shower-to-shower fluctuations in $X_{\mathrm{max}}$ exist even for air showers initiated by the same primary particle, the overall $X_{\mathrm{max}}$ distributions are indicative of the primary mass. Specifically, as the mass of the primary nucleus increases, the mean $X_{\mathrm{max}}$ decreases and the distribution narrows \citep{kampert2012measurements}. This characteristic allows for the selection of proton-induced air showers using an $X_{\mathrm{max}}$ cut. In this analysis, we utilize the same $X_{\mathrm{max}}$ distributions employed in the $\langle\ln A\rangle$ study, and derive an energy-dependent $X_{\mathrm{max}}$ cut that achieves 90\% proton purity, as shown in Figure \ref{fig:proton_cut}. 

\begin{figure}
    \centering
    \includegraphics[width=0.9\linewidth]{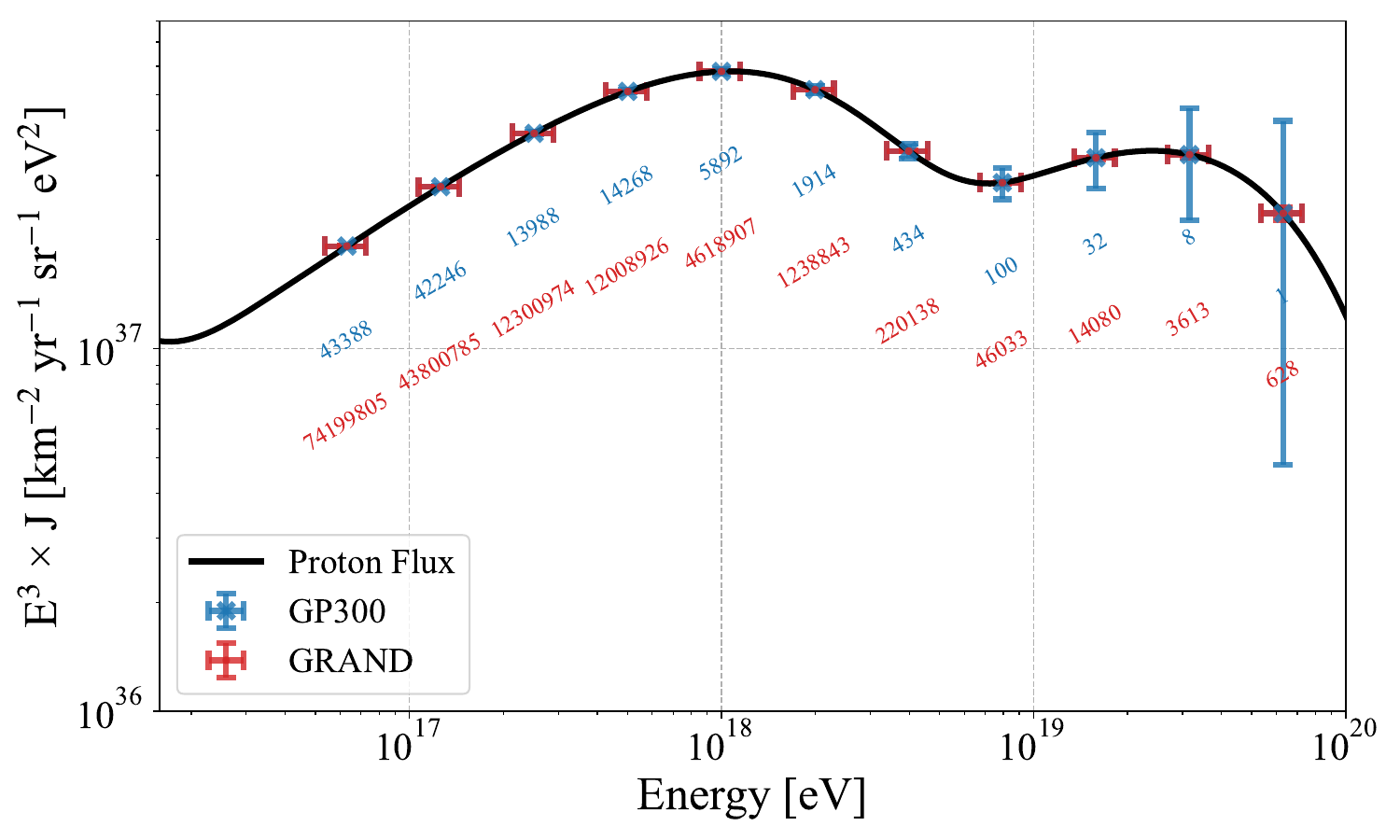}
    \caption{Estimated energy spectrum of CR proton with statistic uncertainties for GP300 (in blue) and GRAND (in red), along with the corresponding event numbers.}
    \label{fig:proton_spectrum}
\end{figure}

By combining this approach with the projected $X_\mathrm{max}$ resolution (e.g., 20 g/cm$^2$) and energy resolution (e.g., 15\%) of the GP300 and GRAND arrays, we estimate the expected proton energy spectrum. The large exposure and sensitivity of GP300 in the 100 PeV–1 EeV range, and of GRAND at the highest energies, ensure that statistically meaningful measurements of the proton component will be achievable across a wide energy range.
Figure \ref{fig:proton_spectrum} illustrates the expected proton spectrum as measured by GP300 and GRAND, both of which achieve reliable resolution within their respective target energy ranges due to their large apertures. 

\subsection{Dipole Anisotropy}

\begin{figure*}[htbp]
    \centering
    \includegraphics[width=0.46\textwidth]{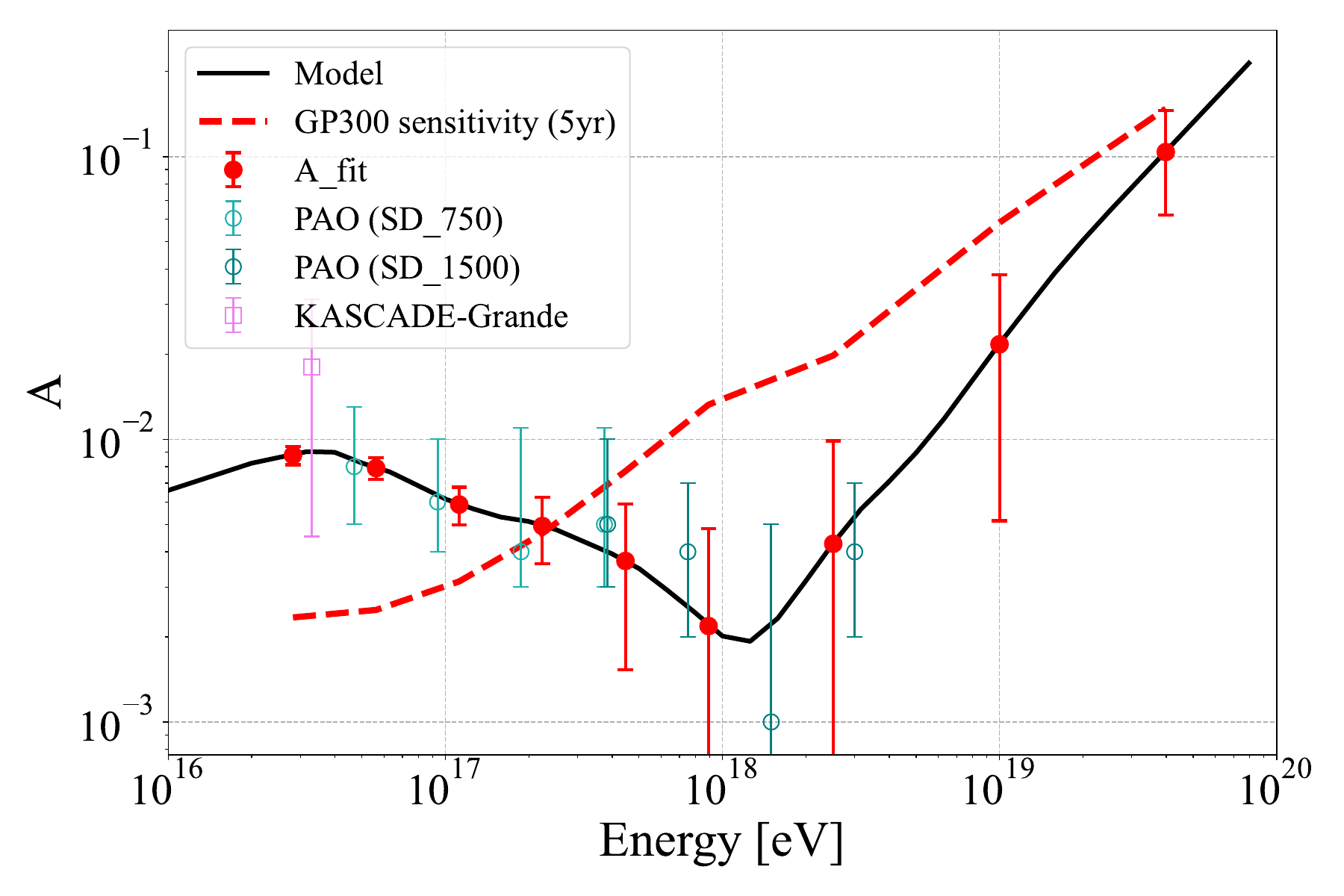}
    \hfill
    \includegraphics[width=0.46\textwidth]{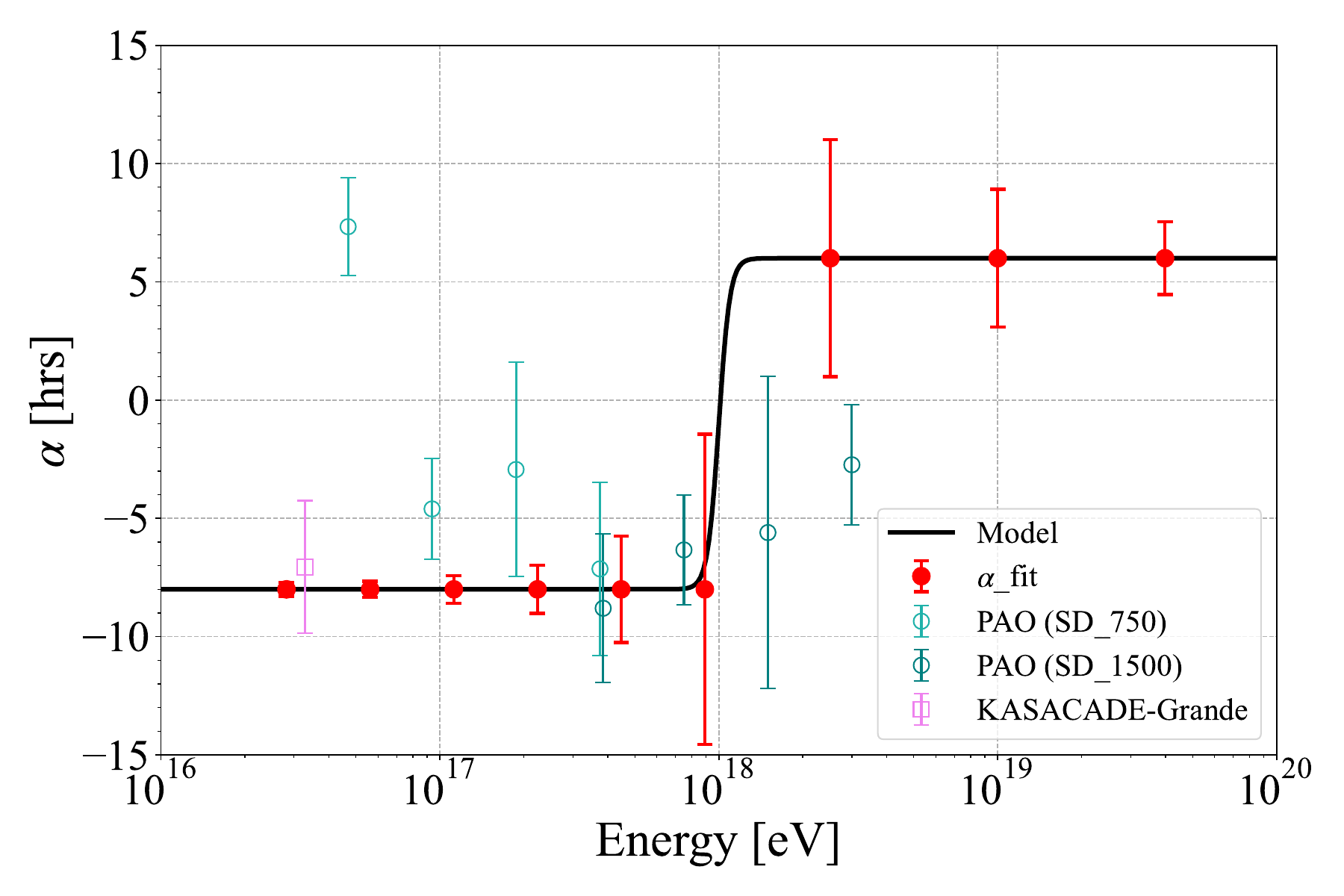}
    
    \vspace{0.5em} 

    \includegraphics[width=0.46\textwidth]{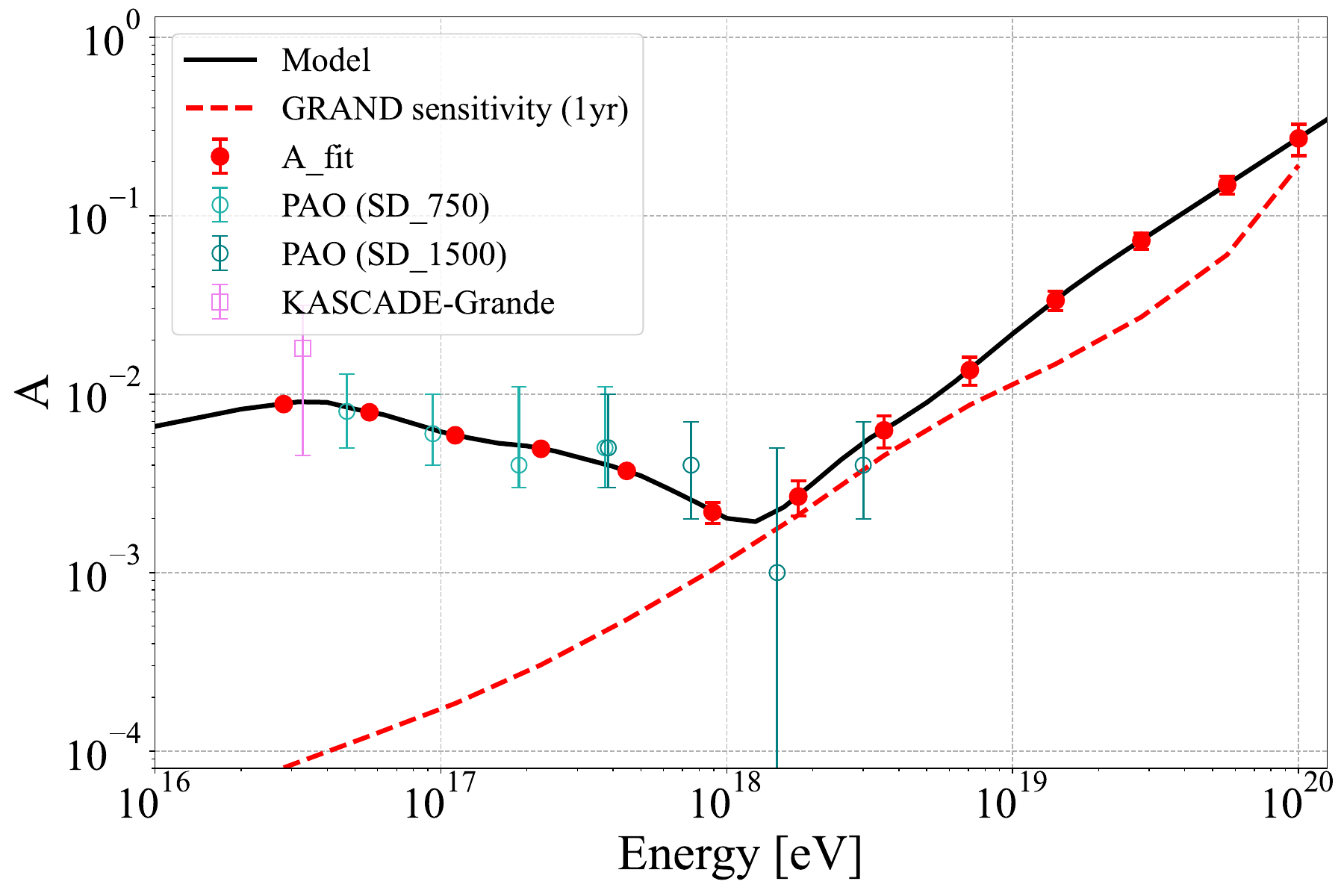}
    \hfill
    \includegraphics[width=0.46\textwidth]{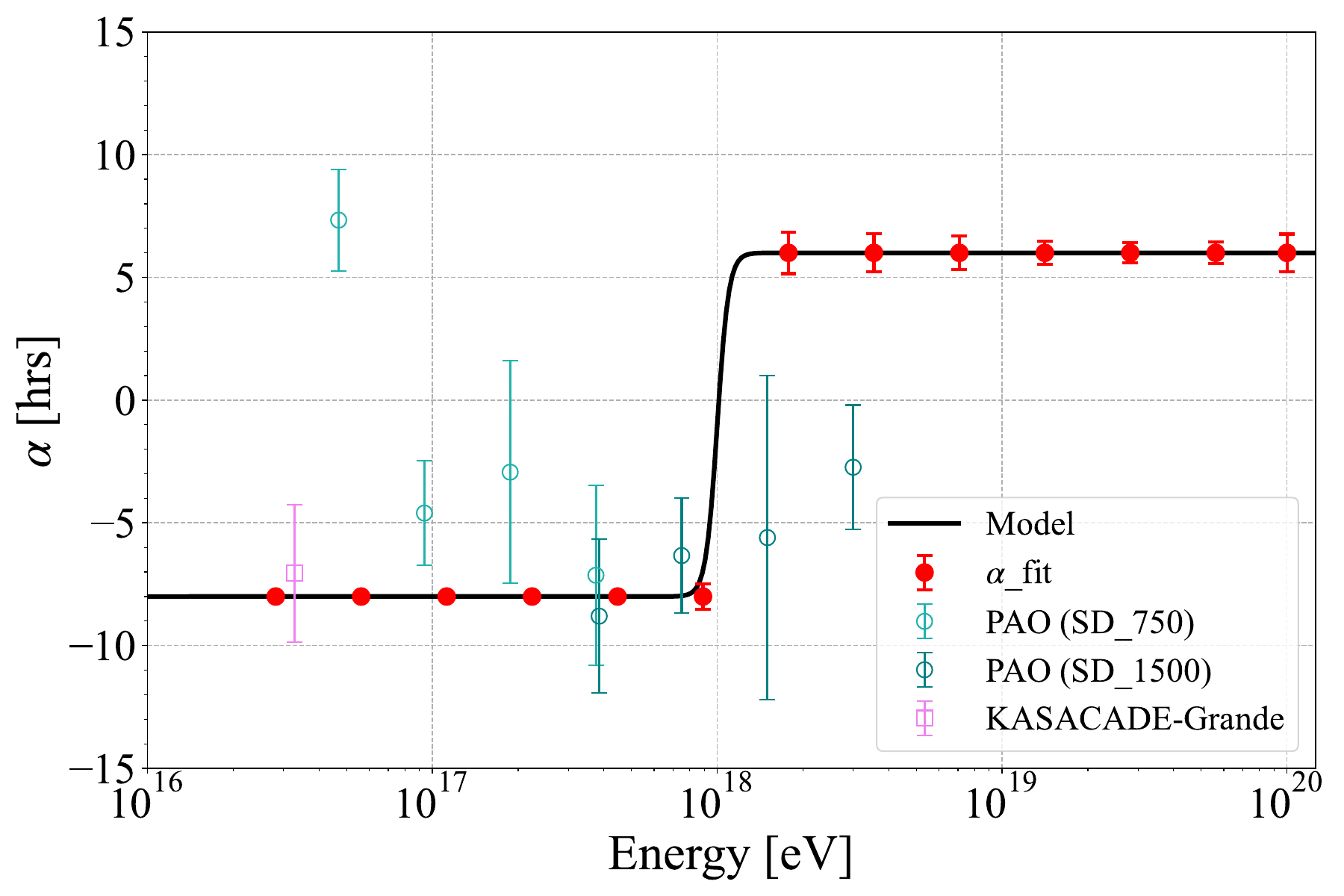}

    \caption{Expected dipole anisotropy measurement sensitivity: top row for GP300 (5 years), bottom row for GRAND (1 year). With comparison of current experimental findings from \citep{KASCADE-Grande:2015xoc, halim2024large}}
    \label{fig:aniso}
\end{figure*}

With the high event statistics anticipated from GP300 and GRAND, measuring large-scale anisotropies in the arrival directions of cosmic rays becomes feasible. The dipole anisotropy can be expressed as:

\begin{equation}
    \frac{\delta I}{\langle I \rangle}=A \cdot \cos (\alpha+\phi)
    \label{eq:aniso}
\end{equation}

where $\langle I \rangle$  is the average cosmic-ray intensity, $\delta I$ is its variation, $\alpha$ is right ascension,  $A$ and $\phi$ represent the dipole amplitude and phase, respectively. In this study, we estimate the dipole anisotropy based on a reference model.

Figure \ref{fig:aniso} illustrates the expected sensitivity of the GP300 array to large-scale anisotropies in the cosmic-ray arrival directions as a function of energy.To accommodate the lower event statistics at higher energies, the energy bin width is increased to 0.6 above $10^{18}$ eV—twice the width used below this energy. The left panel shows the amplitude of the first harmonic in right ascension (denoted $A$), compared with predictions from the reference anisotropy model. The red dashed curve represents the 5$\sigma$ detection threshold after five years of data accumulation. GP300 is expected to detect or constrain anisotropy amplitudes down to the 10$^{-2}$ level at energies around 10$^{17.2}$eV. At higher energies, sensitivity decreases due to the declining cosmic-ray flux. The dots represent the reconstructed amplitudes under nominal experimental conditions, including statistical uncertainties.

The right panel shows the reconstructed dipole phase as a function of energy. To better understand the energy dependence of its uncertainty, the dipole anisotropy can be rewritten as a linear combination of trigonometric components:
\begin{equation}
    \frac{\delta I}{<I>}= A_1 \cos \alpha+A_2 \sin \alpha
\end{equation}

where $A_1=A\cos\phi$, $A_2=-A\sin\phi$. Using standard error propagation, the uncertainty in the phase $\phi$ is given by: 
\begin{equation}
\sigma_{\phi}^2=\frac{1}{A^4}(A_1^2\sigma^2_{A_2}+A_2^2\sigma^2_{A_1})    
\end{equation}
assuming $\sigma^2_{A_1}\approx \sigma^2_{A_2} \approx \sigma^2\propto \frac{1}{N}$,where N is the number of events, we find: $\sigma_{\phi} \propto \frac{1}{A\sqrt{N}}$. This relation explains the observed behavior: the phase uncertainty initially increases with energy due to decreasing event statistics, but decreases at higher energies as the dipole amplitude $A$ grows.

This behavior confirms GP300’s potential to resolve both the amplitude and phase of cosmic-ray anisotropies across a wide energy range.

Figure \ref{fig:aniso} shows the projected sensitivity of the GRAND array to large-scale anisotropies in the arrival directions of cosmic rays, complementing the results shown earlier for GP300. Owing to its significantly larger effective area and higher event statistics at ultra-high energies, GRAND achieves superior sensitivity across a broader energy range.

In the left panel, the 5$\sigma$ sensitivity threshold (red dashed line) in one year remains well below the predicted dipole amplitudes from the reference anisotropy model (black line) up to the highest energies. Unlike GP300, which loses sensitivity above 10$^{18}$eV due to the steeply falling flux, GRAND maintains sensitivity to anisotropy amplitudes at or below the 10$^{-1}$ level even above 10$^{19}$eV, where the event rate becomes extremely low. The fitted amplitudes (red dots) closely follow the model curve, confirming the experiment’s capability to detect or constrain anisotropies with high precision in this regime.

The right panel displays the phase evolution as a function of energy. While both GP300 and GRAND exhibit increasing uncertainty with decreasing amplitude or statistics, GRAND benefits from higher event counts at the highest energies, resulting in improved phase resolution above 10$^{18}$eV. This highlights GRAND’s potential to investigate anisotropy features in a region currently inaccessible to existing experiments.

These results demonstrate that GP300 offers promising sensitivity to detect large-scale anisotropy around 100 PeV, while GRAND extends this capability to the highest energies of the cosmic-ray spectrum.

\section{Conclusion}

Accurate measurements of cosmic rays above 100 PeV are essential to probe the transition from Galactic to extragalactic origins. The radio detection technique, which is sensitive to the electromagnetic component of air showers, provides an independent and complementary approach to particle-based detectors, enabling valuable cross-checks and reducing systematic uncertainties.

The GP300 array targets the 100 PeV–1 EeV range, offering an energy resolution of ~15\% and 20g/cm$^{2}$ $X_{\mathrm{max}}$ precision sufficient for estimating $<\mathrm{ln} A>$ and extracting a high-purity proton spectrum. It is expected to collect $\sim 10^5$ events near 100 PeV annually, providing robust statistics for composition and spectral studies. GP300 is expected to reach sensitivity to large-scale anisotropies down to the level of $5 \times 10^{-3}$ for energies between $10^{16.5} - 10^{17.1}$ eV, enabling not only the detection of dipole amplitudes and phase with meaningful resolution predicted by current models.

GRAND extends these capabilities to higher energies with significantly larger exposure. It maintains strong statistical power well beyond 1 EeV, supporting high-precision measurements of the energy spectrum, mass composition, and anisotropy up to the suppression region. At energies above 10$^{19}$ eV GRAND achieves sensitivity to anisotropy amplitudes at or below 10$^{21}$ and preserves meaningful phase resolution, highlighting its potential to explore directional features in the ultra-high-energy regime. 

Together, GP300 and GRAND are well suited for high-precision radio-based measurements across a wide energy range, offering critical contributions to ultra-high-energy cosmic-ray research.

\begin{acknowledgments}
This work is supported by the National Natural Science Foundation of China (Nos. 12273114), the Project for Young Scientists in Basic Research of Chinese Academy of Sciences (No. YSBR-061) and the Program for Innovative Talents and Entrepreneur in Jiangsu.
\end{acknowledgments}

\bibliography{biblio}{}
\bibliographystyle{aasjournalv7}



\end{document}